\documentclass[twocolumn,showpacs,amsfonts]{revtex4}
\usepackage{amsmath}
\usepackage{latexsym}
\usepackage{float}
\usepackage{amssymb}
\usepackage{graphicx}
\usepackage{textcomp}
\usepackage{hyperref}
\def\ba{\begin{eqnarray}}
\def\ea{\end{eqnarray}}
\def\be{\begin{equation}}
\def\ee{\end{equation}}

\def\bm{\begin{math}}
\def\me{\end{math}}

\newcommand{\dummy}

\begin{document}

\title {Universality in Fluid Domain Coarsening: The case of vapor-liquid transition}
\author{Suman Majumder and Subir K. Das$^{*}$}
\affiliation{Theoretical Sciences Unit, Jawaharlal Nehru Centre 
for Advanced Scientific Research, Jakkur P.O, Bangalore 560064, India}

\date{\today}
\begin{abstract}
~Domain growth during the kinetics of phase separation is studied following vapor-liquid 
transition in a single component Lennard-Jones fluid. Results are analyzed after appropriately 
mapping the continuum snapshots obtained from extensive molecular dynamics simulations to 
a simple cubic lattice. For near critical quench interconnected domain 
morphology is observed. A brief  period of slow diffusive growth is followed by a linear viscous 
hydrodynamic growth that lasts for an extended period of time. This result is in contradiction with 
earlier inclusive reports of late time growth exponent $1/2$ that questions the uniqueness 
of the non-equilibrium universality for liquid-liquid and vapor-liquid transitions.
\end{abstract}
\maketitle
\section{Introduction}
Understanding non-equilibrium evolution during the phase separation in a system is of 
fundamental importance \cite{BRAY,HAASEN,WADHAWAN,JONES} and of much research interest, both 
theoretically and experimentally \cite{HORBACH,DAS,DAS_PURI,MITCHELL,BUCIOR,
YELASH,HORE,MAJUMDER,KENDON,THAKRE,KABREDE,DAS2,BLONDIAUS,LIU,AHMAD,SICILIA,KHANNA,ZANNETTI}. 
Upon quenching from a high temperature homogeneous state to a temperature below the critical point, 
the system becomes unstable to fluctuations and starts phase separating with the formation 
and usually nonlinear growth of particle rich and particle depleted domains. Such 
coarsening is a scaling phenomenon \cite{STAUFFER}, e.g., the shape functions characterizing 
the morphology obey the scaling relations \cite{BRAY,HAASEN,WADHAWAN,SICILIA,STAUFFER}
\begin{eqnarray}\label{scaledCr}
C(r,t) &\equiv& \tilde {C}(r/\ell(t)),\\
S(k,t) &\equiv&\ell(t)^{d} \tilde{S}(k\ell(t)),\\ \label{scaledSk}
P(\ell_{d},t) &\equiv& \ell(t)^{-1}\tilde {P}[\ell_{d}/\ell(t)],\label{scaledPl} 
\end{eqnarray}
where $C(r,t)$, $S(k,t)$ and $P(\ell_{d},t)$ are respectively the two-point equal-time 
correlation function, structure factor and domain size distribution function. 
In Eqs. (\ref{scaledCr}-\ref{scaledPl}) $\tilde {C}(x)$, $\tilde {S}(y)$ and $\tilde {P}(z)$ 
are the master functions independent of time ($t$)-dependent average domain-size $\ell(t)$ that 
typically grows in a power law manner
\begin{eqnarray}\label{powerlaw}
 \ell(t)\sim t^{\alpha}.
\end{eqnarray}
\par
~The growth exponent $\alpha$, in Eq. (\ref{powerlaw}), depends upon the transport mechanism 
driving the phase separation. For diffusive dynamics, associating the domain growth with the 
chemical potential gradient as
\begin{eqnarray}\label{lsdldt}
\frac{d\ell(t)}{dt} \sim \lvert  \overrightarrow{\nabla} 
\mu \lvert \sim \frac{\gamma}{\ell(t)^{2}},
\end{eqnarray}
$\gamma$ being the inter-facial tension, one obtains  $\alpha=1/3$. This is 
referred to as the Lifshitz-Slyozov \cite{LIFSHITZ} (LS) law.
While LS law is the only expected behavior for phase separating solid mixtures, one expects much faster
growth, at large length scales, for fluids and polymers, where hydrodynamics is important. Compared
 to the kinetics in vapor-liquid phase separation, domain coarsening in binary fluids received much more 
attention \cite{HORBACH,DAS,DAS_PURI,HORE,KENDON,THAKRE,AHMAD,SHINOZAKI,DUNWEG,MA,TOXVAERD} 
where following consensus for the behavior of late stage growth \cite{SIGGIA,FURUKAWA,FURUKAWA1,ONUKI} 
has been reached in $d=3$. Considering a balance between the surface 
energy density $(\gamma/\ell(t))$ and the viscous stress ($6\pi\eta v_\ell/\ell(t)$, $v_{\ell}$ being 
the interface velocity and $\eta$ the shear viscosity) for an interconnected domain structure, one can write 
\begin{eqnarray}\label{dldt}
\frac{d\ell(t)}{dt}=v_\ell=\frac{\gamma}{\eta}.
\end{eqnarray}
Solution of Eq. (\ref{dldt}) 
predicts a linear growth ($\alpha=1$), a picture that holds for low Reynolds number. However, 
for $\ell(t)\gg\ell_{in}(=\eta^2/(\rho\gamma)$, $\rho$ being the density), the inertial length, 
the surface energy density is balanced by kinetic energy density $\rho v_\ell^2$, so that
\begin{eqnarray}\label{dlt}
 \frac{d\ell(t)}{dt}\sim \frac{1}{\ell(t)^{1/2}},
\end{eqnarray}
giving $\alpha=2/3$. Note that $\alpha=1$ is referred to as the viscous hydrodynamic growth and 
$\alpha=2/3$ as inertial hydrodynamic growth. While crossover from diffusive to viscous hydrodynamic 
regime was observed in molecular dynamics (MD) simulations \cite{AHMAD,TOXVAERD} and experiments 
\cite{CHOU,WONG,BATES}, both viscous and inertial growths were observed in 
lattice-Boltzmann simulations \cite{WADHAWAN}.
\par
~While non-equilibrium universality for vapor-liquid phase separation is expected to be the same as the 
liquid-liquid, rare inclusive MD simulations that exist \cite{KABREDE,ABRAHAM,KOCH,YAMAMOTO,NAKANISHI} 
for the vapor-liquid transition report an exponent $\alpha=1/2$ for late time domain coarsening. 
Thus, even though our recent focus has turned to systems with realistic 
interactions and physical boundary conditions \cite{HORBACH,DAS,DAS_PURI,MITCHELL,BUCIOR,YELASH,HORE}, 
our basic knowledge and understanding of segregation kinetics 
in bulk fluids still remains a challenge. In this letter, we present results from large scale 
MD simulations for kinetics of vapor-liquid phase separation in a single-component Lennard-Jones (LJ) fluid. 
This work reports first observation of viscous hydrodynamic growth via MD simulation, thus computationally 
confirming that the non-equilibrium universality class of vapor-liquid and liquid-liquid phase 
separation is indeed the same as in the case of equilibrium critical phenomena for both static 
and dynamic properties.
\begin{figure}[htb]
\centering
\includegraphics*[width=0.4\textwidth]{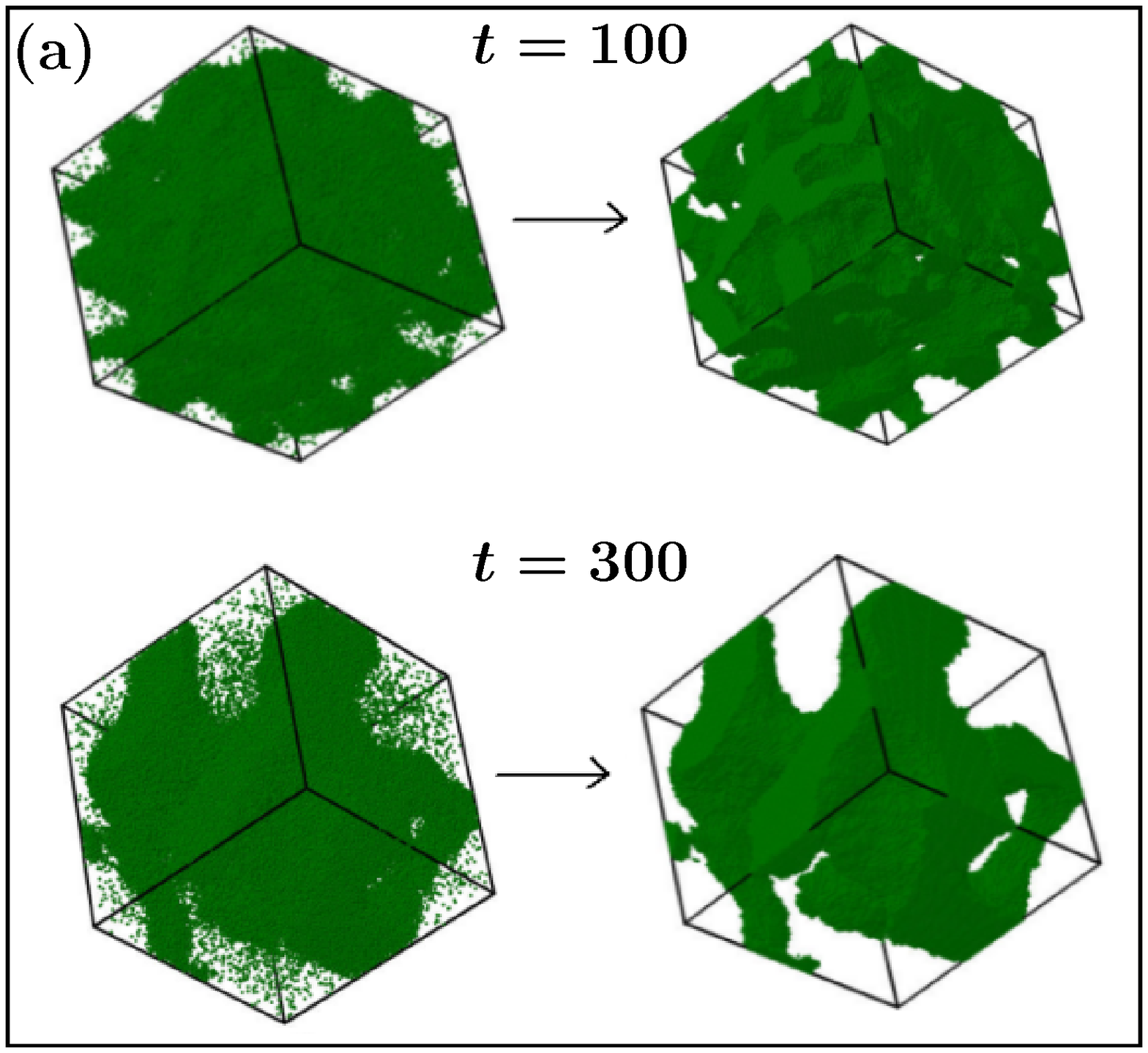}\\
\vskip 0.5cm
\includegraphics*[width=0.4\textwidth]{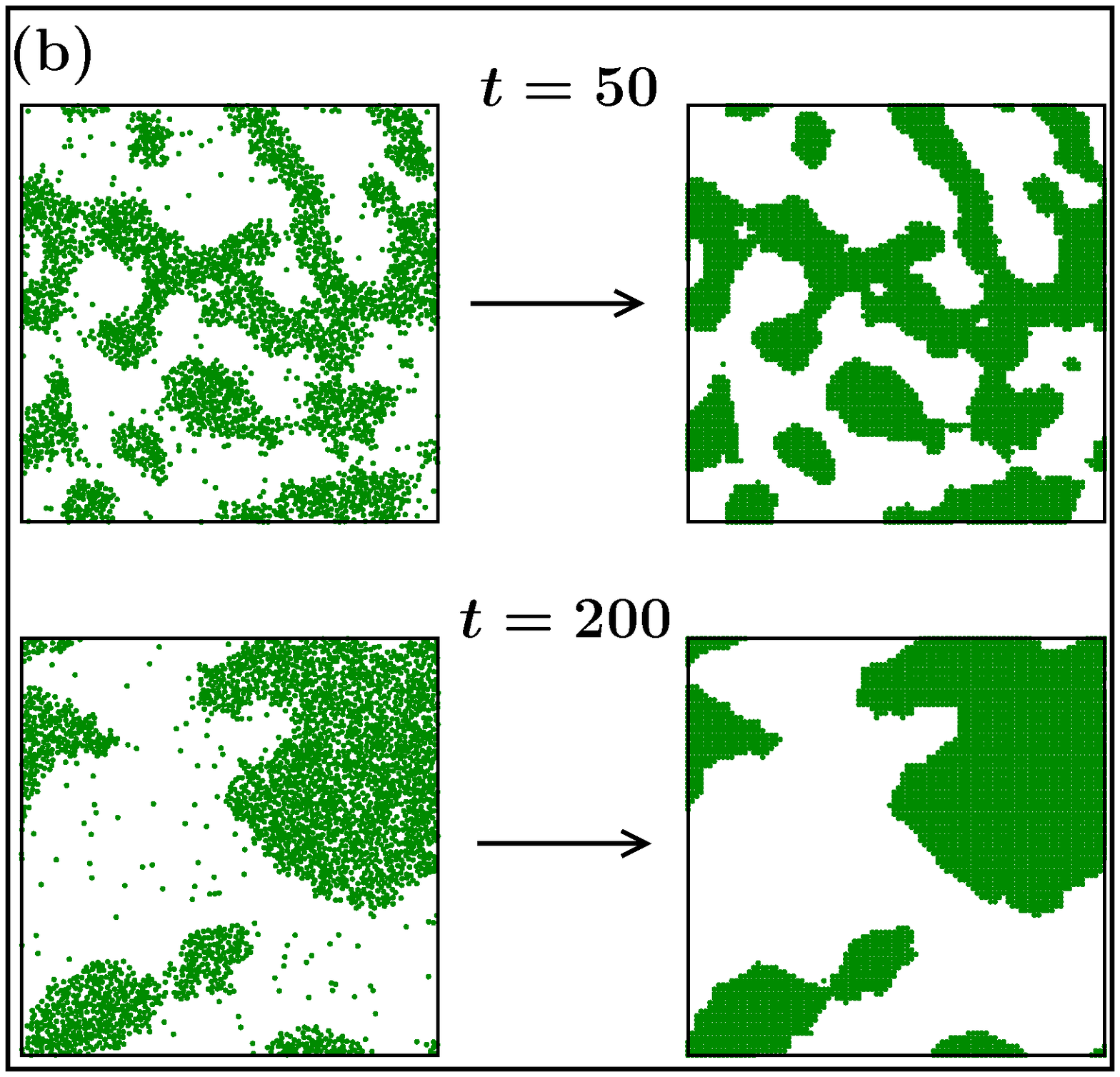}\\
\caption{\label{fig1} (a) Left frames show the $3-d$ snapshots during evolution of the vapor-liquid 
phase separation in a simple Lennard-Jones system, at two different times, as indicated. The total 
density was set to $0.3$ that gives interconnected domain morphology,as seen. 
The simulation was performed after quenching a homogeneous system prepared at a high temperature 
to a temperature $T=0.7$. The right panel corresponds to the same snapshots, but after removing the 
noise (see text for details) to obtain a pure domain morphology. (b) Cross-sectional view of the 
same system in $xy$ plane at two other times. Here also the left frames show the original 
snapshots and the right ones show the same after removing noise.}
\end{figure}
\section{Model and Method}
We consider a model where particles interact via
\begin{eqnarray}\label{vrij}
V(r_{ij}) = U(r_{ij}) - U(r_{c}) - (r_{ij} - r_{c})\left(\frac{dU(r_{ij})}{dr_{ij}}\right)_{r_{ij}=r_{c}},
\end{eqnarray}
with 
\begin{eqnarray}\label{urij}
U(r_{ij})=4\epsilon\left[\left(\frac{\sigma}{r_{ij}}\right)^{12} - \left(\frac{\sigma}{r_{ij}}\right)^{6}\right]
\end{eqnarray}
being the standard LJ potential. In Eqs. (\ref{vrij}) and (\ref{urij}) $r_{ij}$ is the
scalar distance between the $i$th and $j$th particles, $\sigma$ is the particle diameter and
$\epsilon$ is the interaction strength. The cut-off and shifting of the potential at $r_c$ was 
used to faclitate faster computation. Further, introduction of the third term \cite{Allen} on the right
 hand side ensures that both the effective potential and force are continuous at $r_c$ which 
was set to $2.5\sigma$. All particles were assigned equal mass $m$. For the sake of convenience, below we
set $m,~\sigma,~\epsilon$ and $k_B$ to unity. 
\par
~The phase-diagram of the model without the force-correction term is known \cite{Smit}, the critical temperature ($T_c$) 
and critical density ($\rho_c$) being $1.085 \pm 0.005$ and $0.317 \pm 0.006$ respectively. It is understood, of course, 
that the introduction of force-correction will somewhat change the value of $T_c$, which, 
from our experience \cite{Das_Horbach,Das_Fisher,Sutapa} with a similar model that was used 
to study phase-separation in binary mixture, could be approximately $ 10 \%$ lower. 
In this work we have chosen the temperature low enough so that even without 
the complete knowledge of the critical parameters, we are safely below the co-existence curve, as 
is also clear from the nice phase-separating domain pattern shown in Fig. \ref{fig1}. The interconnected 
nature of the morphology also ensures that our choice of overall density ($\rho$) is close to 
the critical value. In a future correspondence, of course, we will undertake the understanding 
of the phase behavior of this model in conjunction with the presentation of detailed results on the 
phase-separation kinetics at various temperatures and densities.

\par
~~A total $265421$ particles were confined in a periodic $3$-dimensional box of linear dimension $L=96$. Thus we 
worked with a density $\rho=0.3$. The MD runs were performed using the Verlet velocity 
algorithm \cite{FRENKEL,Allen} with an integration time step of $\Delta t=0.005\tau$, $\tau=(m\sigma^2/\epsilon)^{1/2}=1$.
The temperature $T$ was controlled by using a Nos\'{e}-Hoover thermostat (NHT) \cite{FRENKEL}, which 
is known to preserve hydrodynamics well. At $t=0$ homogeneous initial configurations, prepared by equilibrating 
the system at $T=5$, were quenched to $T < T_c$. All the results were obtained by 
averaging over $5$ independent runs at a quench temperature $T=0.7$.
\par
~The left frames of Fig. \ref{fig1}(a) are typical $3$-dimensional evolution pictures from 
two different times where the black dots represent a particle. They demonstrate nice interconnected structure 
of segregating domains growing with time. The average density of liquid and vapor domains are 
respectively $0.72$ and $0.02$. A direct method to characterize the structure is to calculate the 
radial distribution function, $g(r)$, from these continuum configurations. However, we are interested in 
looking at $C(r,t)$, its Fourier transform $S(k,t)$ and $P(\ell_d,t)$ which have been more standard tools 
in the study of phase ordering dynamics. To facilitate the calculation of latter quantities, we map the 
continuum snapshots to a lattice one in such a way that a pure domain structure is obtained where 
all lattice sites inside a liquid domain have occupancy one while sites inside the vapor domain have 
occupancy zero. An appropriate renormalization procedure \cite{MAJUMDER,DAS2,AHMAD} is applied for this 
noise removal exercise. 
We further map this system into spin-$1/2$ Ising system where an occupied lattice site has spin value 
$S_i=1$ and for a vacant site $S_i=-1$. The right frames in Fig. \ref{fig1}(a) show such mapped 
configurations. In Fig. \ref{fig1}(b) we demonstrate similar exercise by showing cross-sectional 
views at two other times. Which may be difficult to appreciate from the $3-d$ snapshots, these $2-d$ 
pictures clearly show that no significant structural change has occurred during this transformation that 
can hamper the analysis. Also during the entire duration of study, 
we find that the composition of up and down spins remains conserved within a tolerable limit of $5\%$ 
fluctuations.
\par
~ From the mapped configurations, the correlation function $C(r)$ is calculated as 
\begin{eqnarray}\label{crt}
 C(r,t)=\langle S_iS_j\rangle~-~\langle S_i\rangle^2,
\end{eqnarray}
where the angular brackets stand for statistical averaging. The average domain size $\ell(t)$ from 
the decay of $C(r,t)$ was estimated as 
\begin{eqnarray}\label{lt_crt}
 C(r=\ell(t),t)=h,
\end{eqnarray}
where $h$ was set to the first zero of $C(r,t)$. Other methods employed for the calculation of 
$\ell(t)$ are
\begin{eqnarray}\label{lt_Skt}
\ell(t)=\frac{1}{\int dk~kS(k)}
\end{eqnarray}
and
\begin{eqnarray}\label{lint}
\ell(t)=\int d\ell_d \ell_d P(\ell_d,t).
\end{eqnarray}
In Eqs. (\ref{lt_Skt}) and (\ref{lint}), it is assumed that $S(k,t)$ and $P(\ell_{d},t$) 
are appropriately normalized. In Eq. (\ref{lint}), $\ell_d$ 
is obtained from the separation between two successive domain interfaces in  $x-$, $y-$ or $z-$ directions.
\begin{figure}[htb]
\centering
\includegraphics*[width=0.34\textwidth]{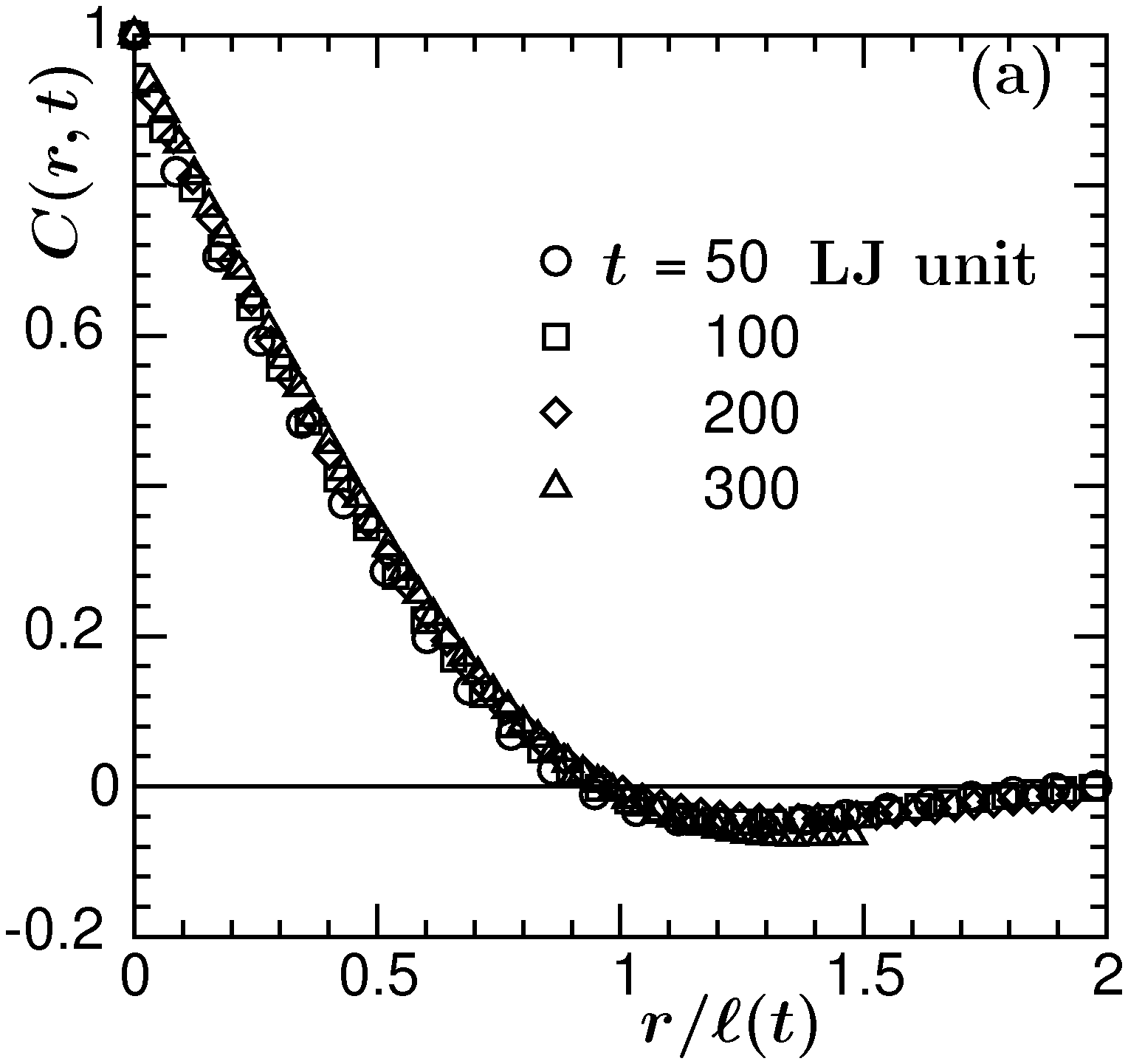}\\
\vskip 0.1cm
\includegraphics*[width=0.34\textwidth]{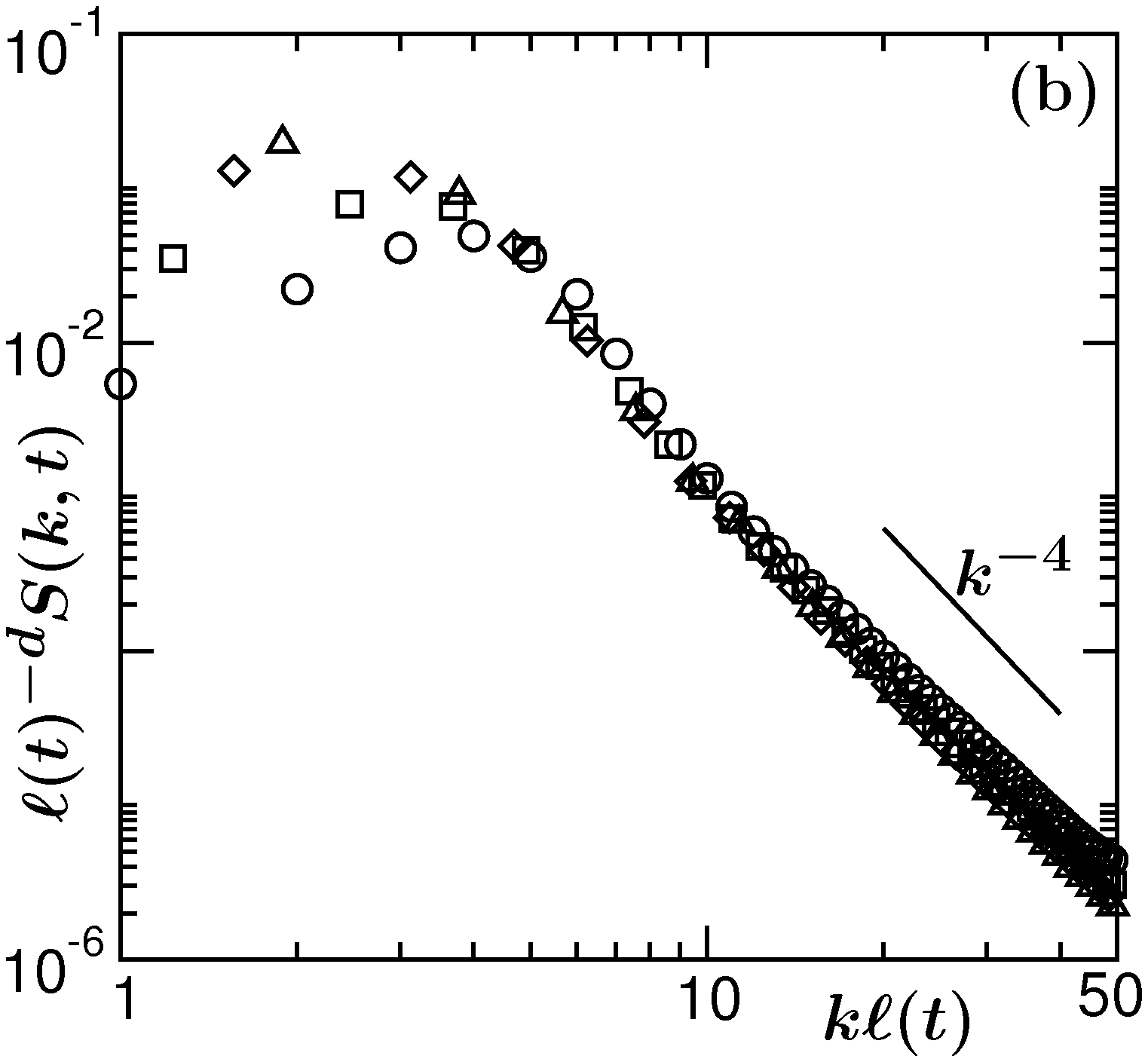}\\
\vskip 0.1cm
\includegraphics*[width=0.34\textwidth]{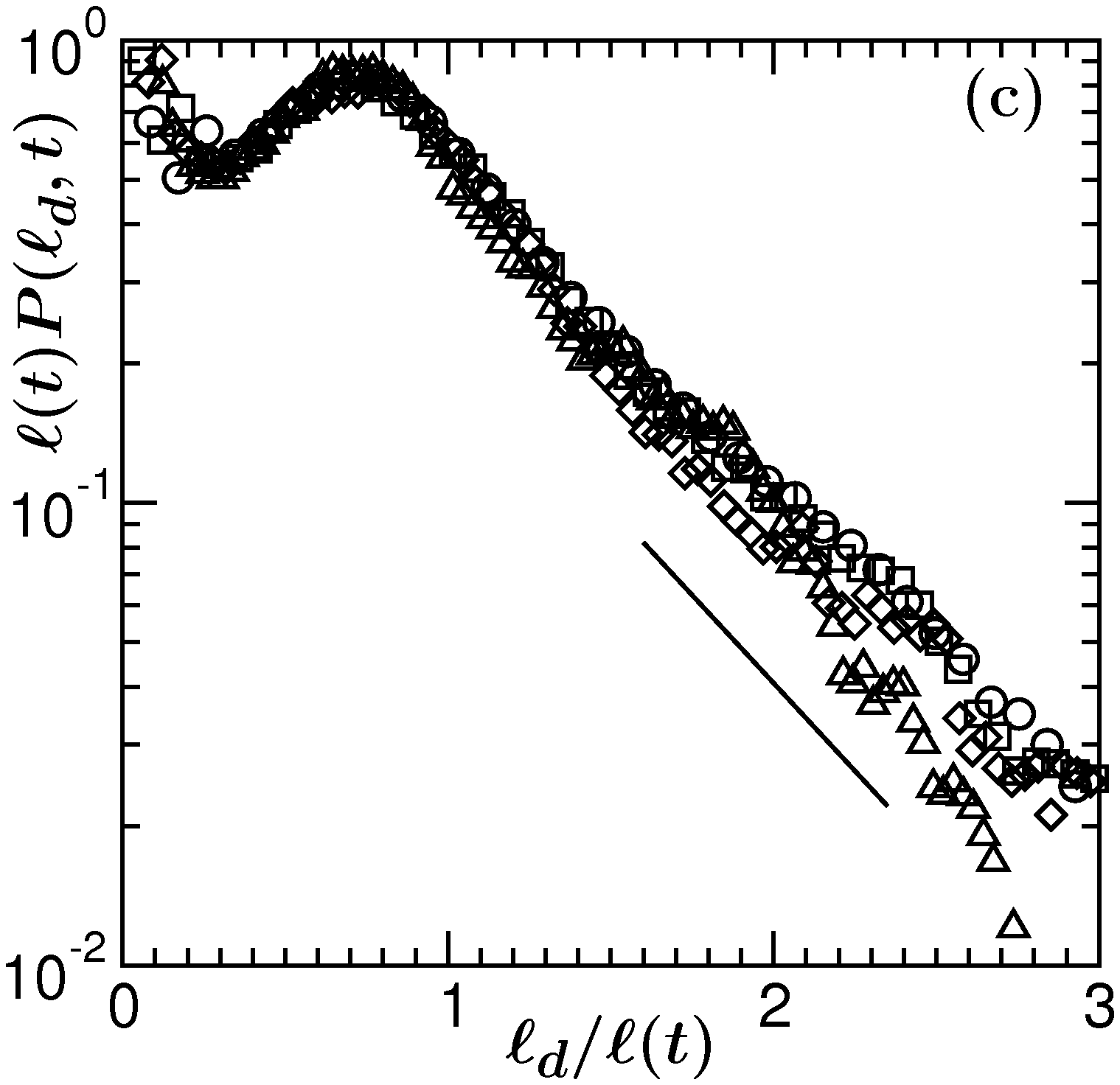}\\
\caption{\label{fig2} Scaling plot of (a) $C(r,t)$ vs $r/\ell(t)$ (b) $\ell(t)^{-d}S(k,t)$ 
vs $k\ell(t)$ and (c) $\ell(t)P(\ell_{d},t)$ vs $\ell_{d}/\ell(t)$ where data from four 
different times, as indicated, have been collapsed. All quantities were calculated from 
noise-free domain structures. The continuous line in (b) corresponds to the Porod law 
while the one in (c) represents an exponential decay of the tail.}
\end{figure}
\section{Results}
In Fig. \ref{fig2}(a), we present the scaling plot of $C(r,t)$ vs. $r/\ell(t)$ for which four 
different times were chosen, as indicated on the figure. While collapse of data from the 
earliest time with the other ones is not good, excellent data collapse with three later times 
is indicative that a scaling regime is reached. In Fig. \ref{fig2}(b) we show the scaling behavior 
of $S(k,t)$ on a log-log plot. The parallel nature of the tail region to the continuous line confirms 
consistency with the Porod law \cite{POROD,OONO_PURI}. Fig. \ref{fig2}(c) shows the scaling plot of 
$P(\ell_d,t)$ on a semi-log plot where the linear behavior of the tail region is consistent 
with an exponential decay \cite{DAS2,SICILIA}. Note that in all the cases, $\ell(t)$ used was 
obtained from Eq. (\ref{lint}). The nice data collapse obtained in all the quantities 
by using $\ell(t)$ from a single measure speaks about the consistency of different methods. 
Henceforth all the results for $\ell(t)$ will be presented after calculating via Eq. (\ref{lint}).
\begin{figure}[htb]
\centering
\includegraphics*[width=0.43\textwidth]{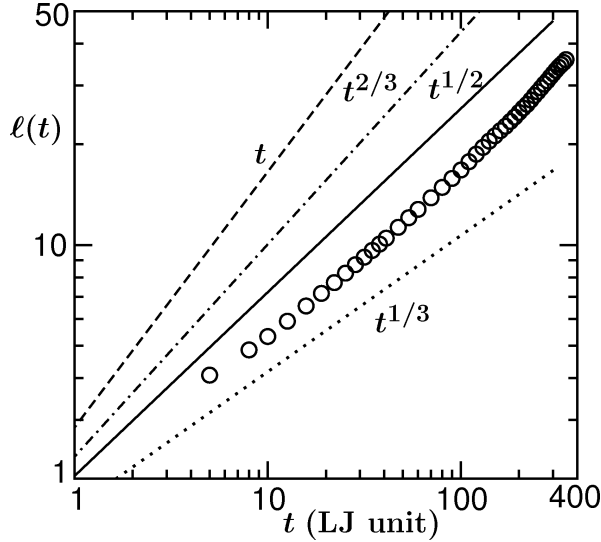}
\caption{\label{fig3} Plot of $\ell(t)$ vs $t$ on a log-scale. All data were obtained 
after removing the noise as described in the text. Here different lines correspond to growth 
laws $\ell(t)\sim t^{1/3},~t^{1/2},~t^{2/3}$ and $t$, as indicated.} 
\end{figure}
\par
~Coming to the central objective of the paper, viz., quantifying the dynamics, in 
Fig. \ref{fig3} we present plot of $\ell(t)$ vs. $t$ on a log scale. There various 
lines correspond to various growth laws in domain coarsening problem. 
While the early time data is more consistent with a diffusive growth ($\alpha=1/3$), 
there appears a gradual crossover to a faster growth. The intermediate time data, of course, 
look consistent with $t^{1/2}$ as reported in previous works \cite{KABREDE,ABRAHAM,KOCH,YAMAMOTO,NAKANISHI} 
and at a later time there seems to be a further crossover. However, due to large offset at 
the crossover region(s), which will be clearer from further analysis and discussions, we 
warn the reader not to take this conclusion seriously. For similar reasons, an apparently 
$t^{1/3}$ looking behavior during early time should also not be taken seriously. In fact, 
due to this and other practical difficulties, it is advisable not to reach a firm conclusion 
from looking at the behavior on a log-scale (which, of course, is a standard practice followed 
in the literature), unless one has results over several decades in time and length.
\begin{figure}[htb]
\centering
\includegraphics*[width=0.43\textwidth]{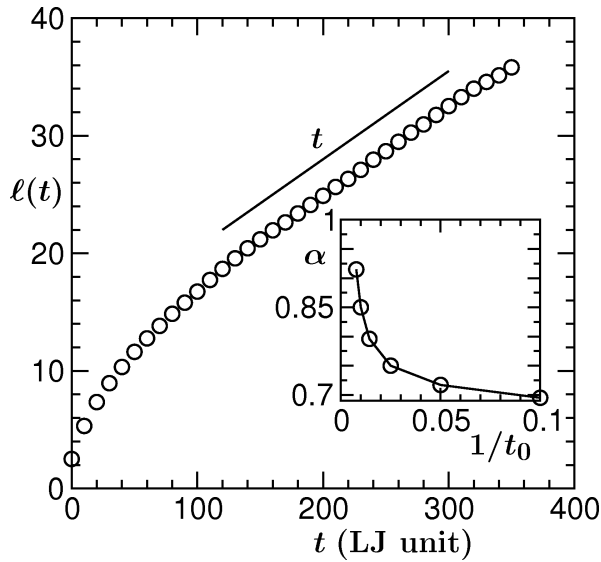}
\caption{\label{fig4} Plot of $\ell(t)$ vs $t$ on a linear scale. All data were obtained 
after removing the noise as described in the text. The solid line corresponds to the expected 
linear growth in the viscous hydrodynamic regime. Inset: Plot of the growth exponent $\alpha$, as 
a function of $1/t_0$, obtained by fitting the data in (a) to the form (\ref{fitting}) in the 
range [$t_0$,$350$].} 
\end{figure}
\par
~Fig. \ref{fig4} shows a plot of $\ell(t)$ vs $t$ on a linear scale where the data starting from $t\simeq100$ 
look quite linear all the way till the end. The deviation from this linear behavior at early time ($t < 100$) could be 
attributed to slower diffusive growth, as discussed above. A fitting to the form, 
\begin{eqnarray}\label{fitting}
 \ell(t)=A+Bt^\alpha,
\end{eqnarray}
however, in the range $t~\epsilon~[0,100]$ gives $\alpha=0.66$. Since this diffusive regime is short lived and also 
accompanied by gradual crossover to a faster growth regime, it is in fact not practical to search for 
growth exponent $\alpha=1/3$ in this region.  On the other hand, a similar fitting to the range 
 $t~\epsilon~[100,350]$ gives $\alpha \simeq 0.9$, quite consistent with the predictions for viscous 
hydrodynamic growth. Note that for $t>350$ one hits the finite-size effects for the system size used in this work. 
 In the inset of Fig. \ref{fig4} we present plot of the exponent $\alpha$ as a function of 
$1/t_0$ obtained by fitting the data to the form (\ref{fitting}) in the range [$t_0$,$350$]. 
This exercise conclusively says that the exponent at late time certainly is larger than $1/2$ which was reported earlier. 
Even though in the main frame of Fig. \ref{fig4} one sees a linear behavior for an extended period of time, 
for the sake of completeness and to further strengthen our claim we take the following route which, 
in addition to being a nice exercise, could be useful for other complex situations.

\par
\begin{figure}[htb]
\centering
\includegraphics*[width=0.43\textwidth]{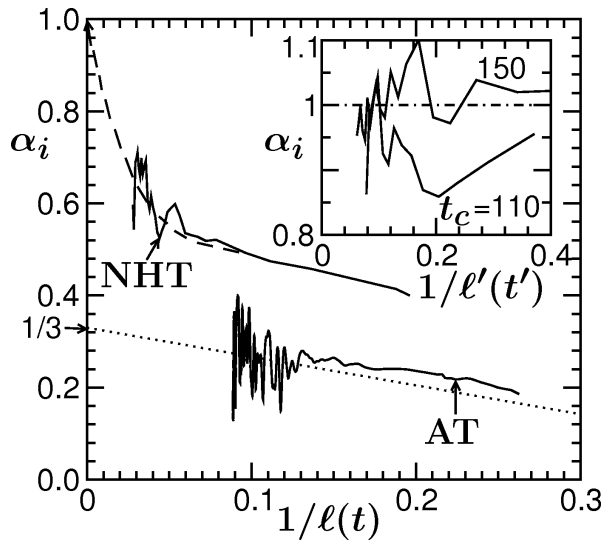}\\
\caption{\label{fig5} Plot of instantaneous exponent  $\alpha_i$ vs $1/\ell(t)$. The dashed 
line is a guide to the eyes, while the dotted line corresponds to Eq. (\ref{alpha_i_2}). 
Inset: Plot of $\alpha_i$ vs $1/\ell'(t')$ for two different values of $t_c$, as indicated. 
 There the horizontal line represents there the linear hydrodynamic growth.} 
\end{figure}

~In Fig. \ref{fig5} we present a plot of instantaneous exponent $\alpha_i$ calculated as \cite{HUSE} 
\begin{eqnarray}\label{alpha_i}
 \alpha_i=\frac{d[\ln \ell(t)]}{d[\ln t]},
\end{eqnarray}
vs $1/\ell(t)$. Due to the off-set ($\ell_0$) at $t=0$, as discussed in the context of Fig. \ref{fig3}, 
$\alpha_i$ is less than unity for the whole range. However it seems to be approaching unity 
in a non-linear fashion. The dashed line there with an arrow at the end serves as a guide to the eyes. The lower 
part in this figure represents the corresponding result obtained from MD simulation by using 
Andersen thermostat (AT) \cite{FRENKEL}. Due to the stochastic nature of this thermalization algorithm, 
it is expected that a diffusive growth will be seen for all time and lengthscales and 
indeed is seen. There the dashed line has the form \cite{MAJUMDER} 
\begin{eqnarray}\label{alpha_i_2}
 \alpha_i=\frac{1}{3}\left[1-\frac{\ell_0}{\ell(t)}\right],
\end{eqnarray}
which says that the observation of $\alpha <1/3$ for finite domain lengths is a mathematical artifact 
coming due to non-zero value of $\ell_0$ and does not necessarily mean that the actual exponent 
is not $1/3$ at early time. Coming to the point of hydrodynamic preserving capability of 
NHT, of course, much better thermostats are available these days. However,
from the difference seen for the results coming from NHT and AT, it is clear that 
NHT is preserving it rather well. Thus the validity of the 
methodology adopted in this work is justified.
\par
~Note that in Eq. (\ref{alpha_i_2}) $\ell_0$ corresponds to the characteristic length scale
at the beginning of a scaling regime and not necessarily the length at $t=0$. 
In situations where one does not expect multiple scaling 
regimes (as is the case with AT) $\ell_0$ can be nicely estimated from a data-collapse 
experiment in finite-size scaling analysis \cite{MAJUMDER,MAJUMDER2}. But situation is more complex 
with NHT due to a crossover to hydrodynamic regime and non-availability of data for different system sizes. 
However, considering that the crossover, at time $t_c$, under discussion is to a linear regime,
confirmation about it could be obtained by using
the following simpler method. To start with, we assume that the growth obeys a power law behavior as a 
function of the shifted time $t'=t-t_c$ as 
\begin{eqnarray}\label{lprime}
 \ell'(t')=\ell(t)-\ell(t_{c})=At'^{\alpha}
\end{eqnarray}
and calculate the instantaneous exponent 
\begin{eqnarray}\label{alpha_prime}
 \alpha_i=\frac{d[\ln \ell'(t')]}{d[\ln t']}.
\end{eqnarray}
Eq. (\ref{lprime}) is invariant under an arbitrary choice of $t_c$ if we are in a 
linear growth regime and one should obtain a constant value $\alpha_i=1$ for all choices of $t_c$ 
in the post crossover regime. In the inset of Fig. \ref{fig5}, we present $\alpha_i$ for two 
different choices of $t_c$. For $t_{c} > 110$, $\alpha_i$ oscillates around the mean value $1$ 
which is consistent with the viscous hydrodynamic regime. This oscillation (which could as well 
be appreciated from main part of Fig. \ref{fig5}) is also observed 
in other studies \cite{SHINOZAKI,MAJUMDER}. In a finite system, at later time only a few domains of 
comparable size exist which are separated from each other by large distances. Thus, as time 
increases, it takes longer time for them to merge which causes the exponent to come down. 
Finally merging of two huge domains after a long interval suddenly enhances the value 
of the exponent.
\vskip 1cm
\section{Conclusion}
We have presented results for the kinetics of vapor-liquid phase-separation from 
the molecular dynamics (MD) simulation of a simple Lennard-Jones fluid using both Andersen (AT) and
Nos\'{e}-Hoover (NHT) thermostats. It is observed that NHT is reasonably useful for studying 
hydrodynamic effects in the fluid phase separation. A brief period of diffusive coarsening was followed 
by a linear viscous hydrodynamic growth. Our results are in contradiction with few previous 
MD studies which reported an exponent much less than unity, however, is consistent with 
the results of binary fluid phase separation. One requires much larger system size to observe 
the inertial hydrodynamic growth. This we leave out for a future exercise, in addition to
the study of coarsening of liquid droplets for off-critical quench. For the latter problem one 
expects single asymptotic exponent $1/3$ from droplet diffusion-coagulation mechanism \cite{STAUFFER,BINDER2}. 
Also, as discussed, MD with AT and a Monte Carlo simulation of the same system will provide 
a single diffusive $1/3$ growth for all compositions. For off-critical composition the latter 
exercise should be able to provide information on the curvature dependent inter-facial tension 
which is expected to bring in early-time correction in the LS law. This will 
certainly be interesting to compare with the corresponding results obtained from 
equilibrium studies \cite{Benjamin}.
\acknowledgments 
SKD acknowledges discussions with S. Puri and S. Ahmad. The authors acknowledge grant 
number SR/S2/RJN-$13/2009$ of the Department of Science and Technology, India. 
SM is also grateful to Council of Scientific and Industrial Research, India, for financial support.
\par
\hspace{0.2cm}${*}$ das@jncasr.ac.in

\end{document}